\documentclass[sts]{imsart}

\RequirePackage{amsthm,amsmath,amsfonts,amssymb}
\RequirePackage[numbers]{natbib}
\RequirePackage[colorlinks,citecolor=blue,urlcolor=blue]{hyperref}
\RequirePackage{graphicx}
\RequirePackage{tikz}
\usetikzlibrary{positioning,arrows.meta,shapes}

\startlocaldefs
\numberwithin{equation}{section}

\theoremstyle{plain}

\theoremstyle{definition}

\endlocaldefs

\begin{document}

\begin{frontmatter}

\title{Comment on: ``The Two Cultures of Prevalence Mapping: Small Area Estimation and Model-Based Geostatistics''}

\begin{aug}
\author[A]{\fnms{Emanuele} \snm{Giorgi$^{1,2}$}\ead[label=e1]{e.giorgi@bham.ac.uk}}
,
\author[A]{\fnms{Claudio} \snm{Fronterre$^{1}$}\ead[label=e2]{c.fronterre@bham.ac.uk}}
\and
\author[B]{\fnms{Peter J} \snm{Diggle$^{1,2}$}\ead[label=e3]{p.diggle@lancaster.ac.uk}}

\address[A]{$^1$Department of Applied Health Sciences, University of Birmingham, Birmingham, B15 2TT, United Kingdom\printead[presep={, }]{e1,e2}}

\address[B]{$^2$Lancaster Medical School, Lancaster University, Lancaster, LA1 4YW, United Kingdom\printead[presep={, }]{e3}}

\end{aug}



\end{frontmatter}

\section{Introduction}

Wakefield and colleagues \citep{wakefield2025two} have produced what is likely to become an essential reference in the field of prevalence mapping by articulating a much-needed synthesis of Small Area Estimation (SAE) and Model-Based Geostatistics (MBG) approaches. We agree with their central thesis that practitioners should avoid dogmatic adherence to either tradition and instead select methods appropriate to the specific inferential goals and data constraints at hand. This represents a pragmatic approach to statistical practice that draws judiciously from multiple methodological tools while maintaining healthy skepticism about the assumptions and limitations inherent in each approach, rather than treating any single framework as methodologically sacrosanct.

 Wakefield and colleagues are commendably thorough in their exposition of SAE methods but, in our opinion,
 there are some missing aspects in their treatment of MBG, which is a more nuanced modelling framework than is described in their paper.  We hasten to add that this is not negligence on the authors' part, but rather stems from two understandable constraints: firstly, the focus of their paper is predominantly on national cross-sectional surveys, such as the Demographic and Health Surveys (DHS), that employ well-documented stratified cluster sampling designs; secondly, the comparison between the two frameworks is restricted to the goal of producing area-level predictions. In these circumstances, and with one caveat, we would agree that SAE methods constitute a natural approach, as is reflected in their widespread use for the construction of health atlases; see, for example
 \cite{hansell2014} or \cite{ramis}.
 The caveat is that SAE methods handle spatial dependence by imposing a neighbourhood structure
 that, in practice, is  unquestioned but is built from
 very simple considerations. The most widely used neighbourhood structure is that two small areas are neighbours if
 and only if their boundaries are contiguous. This is reasonable for
 compact geographies, among which the
 mainland d{\'e}partments of France is a good example, but is less appealng for more convoluted geographies. 
 
In the same anti-dogmatic spirit as Wakefield and colleagues \citep{wakefield2025two}  we would like to draw a more complete picture of MBG for prevalence mapping.
We would argue that a fair comparison of the two modelling philosophies needs to consider settings that extend well beyond the two restrictions listed above. In this spirit, we focus on the following points that we believe merit a fuller
and more nuanced discussion:

\begin{enumerate}
\item the role of model interpretability in the model specification;
\item the choice and specification of covariates, including whether design variables should take priority, or
even remain necessary, when a rich
collection of spatial covariates is available;
\item inferential goals that go beyond area-level prediction;
\item handling multiple data sources
recorded on incommensurate partitions
of the study-area;
\item tailoring model validation strategies to the specific inferential task.
\end{enumerate}
Addressing these points requires 
perspectives that extend beyond statistical methodology. Substantive expertise in disease epidemiology and the policy context in which the model inferences will be used are essential for making informed modelling choices. Accordingly, 
in what follows we elaborate on each of these points and argue that in many cases the choice between SAE and MBG approaches should be informed by considerations that go well beyond formal validation metrics. 


\section{Design ignorability in model-based Geostatistics}

Wakefield and colleagues \cite{wakefield2025two} state that ``MBG approaches typically assume the sampling design is ignorable'' and ``often pay less attention to the sampling design.'' This critique may suggest that MBG is fundamentally incompatible with complex survey designs. On the contrary,  MBG
can include design-based adjustments  and, more generally, can achieve design ignorability by conditioning on design variables and/or scientifically motivated covariates. 

A more insidious problem with MBG is the potential for preferential sampling, which arises when there is stochastic dependence between the location process, $X$  say, and the outcome $Y$ associated with each of the sampled locations \citep{diggle2010}. The analogous problem for SAE is informative missingness, which potentially arises when not every small-area in the study-region supplies data or,less transparently, when the recorded outcomes in some small-areas are based
on incomplete information.

To examine this aspect of MBG more formally, we write $[\cdot]$ for ``the probability distribution of $\cdot$'' and let $D$ denote the design variables. The joint distribution of sampling locations $X$, outcomes $Y$, and design variables $D$ can always be expressed as:
\begin{equation}
    \label{eq:jd}
    [X, Y, D] = [D] \times [X \mid D] \times [Y \mid X, D].
\end{equation}
 When $[Y \mid X, D] = [Y \mid D]$ or,
 equivalently,  $Y \perp X \mid D$, the factorization \ref{eq:jd} shows  why conditioning on $D$ is sufficient for design ignorability. The conditional independence, $Y \perp X \mid D$, implies that once we account for the design variables in our model, the sampling locations provide no additional information about the outcomes. The critical insight here is that failing to condition on $D$ collapses equation \eqref{eq:jd} to the marginal relationship $[X, Y] = \int [X, Y, D] \, dD$, which generally does not factorize as $[X] \times [Y]$ unless $D$ has no effect on $Y$. When design variables influence both sampling probabilities and health outcomes, ignoring $D$ creates a stochastic dependence between $X$ and $Y$ which fits precisely the definition of preferential sampling. Any statistical model for the data should thus aim to break the stochastic dependence between $X$ and $Y$, with the introduction of $D$ being the most natural choice according to the authors in \cite{wakefield2025two}. However, as we argue in the next section, our view is that one should seek models that can provide the best possible explanation of the variation in $Y$ and eliminate the stochastic dependence between $X$ and $Y$, using $D$ primarily as a diagnostic to confirm design-ignorability rather than as the modeling solution itself.

\section{Conditioning on mechanistic mediators}

Generally, design variables do not have a direct effect on the outcome of interest. Instead, their association with disease risk operates through a variety of environmental and socio-economic pathways. For example,
in the analysis of Zambia HIV prevalence, the treatment of urban/rural classification as a binary factor  is empirically justifiable because urbanicity is associated with HIV prevalence, but raises the question of whether conditioning on variables more directly related to the outcome—such as those mediating the effect of urbanicity on HIV risk—might be a better strategy.  
More direct measures, depending on the health outcome in question, might include air quality,
population density, demography, access to clean water, etc. 

This observation motivates an alternative formulation in which we condition on variables $C$ that mediate the effect of the design variables $D$ on the outcome $Y$.

Let $C$ denote a set of covariates representing hypothesised pathways through which $D$ influences $Y$. If these covariates fully mediate the effect of $D$ on $Y$, we can express the joint distribution of $Y$,
$X$, 
$C$ and $D$ as:
\begin{equation}
    \label{eq:mediated}
    [Y, X, C, D] = [D] \times [C \mid D] \times [X \mid D] \times [Y \mid C]
\end{equation}
where the key assumption is that $Y \perp D \mid C$ -- that is, conditional on the mediating variables $C$, the design variables $D$ provide no additional information about the outcome. Under this formulation, the conditional distribution of interest becomes:
\begin{equation}
    \label{eq:conditional_mediated}
    [Y \mid X, C, D] = [Y \mid C]
\end{equation}
which shows that once we condition on $C$, neither the sampling locations $X$ nor the design variables $D$ affect the outcome distribution, thus  achieving design ignorability.

Admittedly, this raises two practical challenges. Firstly, we must ensure that our chosen set $C$ captures all pathways from $D$ to $Y$. Incomplete mediation, in which case some residual effect of $D$ remains, would violate the conditional independence assumption and reintroduce residual bias as a consequence of informative sampling. Secondly, the variables in $C$ must be reliably measured for both sampled and unsampled locations to enable valid aggregation to area-level predictions. Wakefield and colleagues acknowledge this possibility when they note that ``readily available variables such as population density and/or nighttime lights'' could serve ``as surrogates for urban/rural,'' but they counsel caution: ``The success of this strategy is difficult to anticipate.'' We would agree that even the most careful attempt to identify all mediating variables will be less than perfect but we would argue that  treating design variables as additive  effects has its own substantive limitations. In the context of classically designed experiments, randomisation delivers valid design-based inference, but this justification does not hold in prevalence mapping applications.

The binary urban/rural dichotomy, although convenient for sampling design, is epidemiologically unsatisfying for several reasons. Urbanicity is inherently continuous and multidimensional, and a binary classification can mask substantial heterogeneity. The dichotomisation hides the gradations in disease risk that may be most epidemiologically relevant. Moreover, as Wakefield and colleagues note, urban/rural classifications ``can be updated between the time of the census and the survey'' and ``may change over time due to urbanization.'' The method they use to create pixel-level urban/rural maps—thresholding population density—introduces arbitrary decisions that propagate into the modeling. Figure S1 in their supplement illustrates this starkly: the vast majority of Zambia is classified as rural, with only sparse urban pixels. While the authors appropriately emphasize the importance of acknowledging sampling design, their treatment reveals an asymmetry: they counsel caution about conditioning on mechanistic mediators while expressing confidence that including design variables is "the safest option." This asymmetry risks instilling excessive confidence that including design variables is sufficient for achieving design ignorability, potentially discouraging exploration of alternative models where rich mechanistic mediators $C$ might render the design variables $D$ redundant.

More fundamentally, the association between urbanicity and disease risk is mediated by disease-specific mechanisms that differ markedly across conditions. For diseases where the mechanistic pathways are well understood and measurable, we would prefer a different modeling strategy than that advocated by Wakefield and colleagues: begin by conditioning on the richest available set of covariates $C$ that are judged by disease experts to be epidemiologically relevant for the outcome, and only as a final diagnostic step—rather than as an a priori requirement—include design variables to test whether their effect has been adequately explained by the set
$C$. This approach reverses the burden of proof: instead of assuming design variables are necessary and testing whether alternatives might suffice, we ask whether, in settings where randomisation is not available, design variables add meaningful information beyond what is captured by disease-specific mechanistic pathways. If the design variables remain statistically significant and improve model fit after conditioning on mechanistic covariates, this reveals incomplete mediation and justifies their inclusion. If they do not, their inclusion may introduce noise without reducing bias. Wakefield and colleagues' own results hint at this possibility: in their Table S2, the urban coefficient attenuates from 0.878 to 0.641 when just three covariates (access, malaria, nighttime lights) are added, suggesting partial mediation even with this limited covariate set. A richer assembly of mechanistic covariates—population density gradients, wealth indices, healthcare access, demographic composition, and disease-specific environmental determinants—may attenuate the design variable effect further, potentially to statistical and practical insignificance.

\section{The wider use of geostatistical models}

In our opinion, SAE methods are a good choice for obtaining area-level estimates from well-designed, geographically comprehensive surveys. But many problems concerning the geographical distribution of health outcomes exhibit  complexities that cannot be accommodated by SAE. 
 
Often, the aim of an analysis is not only to predict at the area-level on which data are assembled. The spatially continuous underpinning of MBG models enables them to be used simultaneously to draw inferences at different spatial scales. Typically, the output of an MBG analysis includes a sample from the predictive distribution of the complete prevalence surface, ${\cal P} = \{P(x): x \in A\}$. Predictive inference about any property of ${\cal P}$ follows automatically, including but not restricted to the joint distribution of prevalence on any partition of $A$. 
 
Similarly, MBG avoids the need for ad hoc solutions to the so-called modifiable areal unit problem (MAUP, \cite{openshaw1984}). A not-uncommon scenario is that relevant covariate information and health outcome data are available on different partitions of the study-region. MBG can handle this by formulating a spatially continuous model and deriving the associated likelihood for the observed health outcome data. Diggle and colleagues \cite{diggle2013} give an example in which covariate information at 100 metres resolution was combined with counts of cancer deaths on an adminstratively defined partition of the Castile-La Mancha region of Spain to identify  variations in cancer at spatial scales smaller than the administrative areas. Johnson and colleagues \cite{johnson2020} give another example of an MBG approach, which they used to model spatially aggregated life expectancy outcomes in the Liverpool, UK council area, and to understand its association with an index of multiple deprivation based on a different administrative partition.

A further complication arises when combining data from multiple surveys with different sampling designs, particularly when design variables are unavailable or when data are available from both randomized and non-randomized surveys. In such settings, the design-based foundations of SAE methods become untenable. Giorgi and colleagues \cite{giorgi2015} demonstrate how geostatistical models can successfully combine data from multiple spatially referenced prevalence surveys by treating all observations through a unified spatial model that borrows strength across surveys through shared spatial and covariate structures. This approach, coupled with the use of mechanistic covariates as described in the previous section, provides a coherent model-based framework for synthesis that does not require complete design information from all contributing surveys.

\section{Model comparisons}
An issue common to both SAE and MBG is how to evaluate predictive performance. Validation approaches should be tailored to the predictive goal, which is typically interpolation within the study area, though in some cases extrapolation beyond it may also be of interest. The leave-one-out cross-validation approach used in Section 6 confounds these distinct scenarios. Holding out peripheral administrative areas effectively evaluates extrapolation performance, while holding out interior areas evaluates interpolation. Since the intended use of these models is interpolation within the study region, evaluating against a criterion that conflates both scenarios may undermine comparison. Second, and more critically, model performance is assessed against the observed direct estimates as if they were ground truth. This issue is particularly acute in low-prevalence settings, where design-based estimators may be unstable. 

A further point concerns the treatment of unstructured overdispersion in prediction. Although the ``nugget effect'' is accounted for at the likelihood level, either through explicit cluster-level random effects or via beta-binomial sampling models, the authors appear to exclude it from the leave-one-out cross-validation predictions. While the nugget may not be of scientific interest, it must be included when the predictive target is the observed outcome itself. Because of the nonlinearity of the logistic link, excluding the nugget can induce downward bias in aggregated prevalence estimates and lead to overly narrow uncertainty intervals. This mechanism offers a plausible explanation for both the downward bias and the undercoverage observed in the Zambia application, suggesting that these findings may reflect implementation and validation choices rather than intrinsic properties of model-based geostatistical approaches.

Beyond this technical issue which can be easily addressed, the leave-one-out cross-validation in Section 6.3 has structural features that complicate interpretation. The comparison is fundamentally asymmetric across model families. When an Admin-2 area is left out for the Fay-Herriot model, one observation—the area-level direct estimate—is removed. For the unit-level models, the same operation removes multiple clusters' worth of individual-level data. The information loss differs in both character and magnitude. Moreover, the Fay-Herriot model has an inherent advantage in this validation setup because its sampling model explicitly specifies that the direct estimate equals the true parameter plus normal noise with known variance. The leave-one-out procedure then asks each model to predict the direct estimate, which is precisely the quantity the Fay-Herriot model is designed to model. Unit-level models, by contrast, are specified at the individual or cluster level and are designed to predict an underlying latent prevalence or risk surface rather than the design-based direct estimate itself. To enable comparison, their predictions must first be aggregated to the area level and then transformed onto the same scale as the direct estimator. Consequently, these models are effectively evaluated against a derived proxy rather than against the latent population quantity that they are intended to infer.

We suggest that future research could focus  on developing simulation-based approaches to model validation, in which the performance of one or more candidate models is assessed against simulated data using metrics that are directly linked to the operational use of the models. Since, for this purpose, the simulation model does not need to be empirically identifiable it can incorporate contextually informed data-generating mechanisms, with no
penalty for complexity. This approach is an overt response to the truism that all models are wrong but some are useful
\citep{box1976}, shifting the emphasis from self-referential correctness to robustness against plausible departures from the assumed model, judged by metrics that more directly reflect the intended use of the model in question; for example, in Neglected Tropical Disease control programmes, the policy-relevant question is often not prevalence in itself, but whether and where local prevalence exceeds an agreed intervention threshold. Such a simulation framework would allow generation of the true classification of each areal unit with respect to the threshold, and different candidate models fitted to the simulated survey data could then be evaluated on their ability to correctly classify areas as above or below the intervention threshold. We believe this provides a more robust alternative to cross-validation and directly assesses model performance in terms of their intended operational use.

\section{Conclusions}

SAE is a principled smoothing method, highly appropriate for spatially discrete data obtained from area-wide surveys
conducted according to a
well-documented sampling framework,
such as the Zambia DHS survey described in
\cite{wakefield2025two}. 

MBG is an equally principled methodology for analysing incompletely observed spatially continuous phenomena based on a hierarchical modelling framework,
$[P,Y]=[Y][Y|P]$ where $[P]$ is the unconditional distribution of the unobserved scientific {\it process},
$[Y|P]$ is the conditional distribution of the {\it data}
given the realisation of the scientific process and the
specification of $[P]$ is informed by contextual knowledge.

 When design variables are available, they can be used equally by SAE or MBG, but alternative models that are built on a judicious selection of covariates can be scientifically more informative, and may even deliver better predictive performance. Leave-one-out cross-validation plays a useful but limited role in assessing predictive performance. Simulation-based methods that assess model performance linked to their primary purpose provide complementary information on model performance and should be further explored.

\bibliographystyle{imsart-nameyear}
\bibliography{references}

\end{document}